# Los Cultivos Mixtos y las Fermentaciones Alcohólicas

Pilar Escalante-Minakata[1] y Vrani Ibarra-Junquera[2*],

[1]*División de Biología Molecular, IPICyT; E-mail: minakata@ipicyt.edu.mx*

[2]*Facultad de Ciencias Químicas, Universidad de Colima, E-mail: vij@ucol.mx*

RESUMEN

La dinámica poblacional de cultivos mixtos y secuenciales en fermentaciones de mostos de frutas es un problema muy interesante desde el punto de vista teórico y tecnológico. Por cultivos mixtos nos referimos a que se encuentran presentes desde un inicio más de una especie/raza de microbio y por cultivo secuencial a que serán añadidas a lo largo de la fermentación. Dichos procesos tienen un papel fundamental en la industria de bebidas fermentadas, de ahí la importancia de su estudio y modelado. Los modelos de estos procesos deben representar la dinámica de múltiples especies de microorganismos que crecen en una mezcla de sustratos, porque el mosto de frutas esta formado por una proporción importante de hexosas y pentosas. En las bebidas y en general en los alimentos fermentados un aspecto muy importante son las propiedades organolépticas, por ello resulta fundamental estudiar la dinámica poblacional y el impacto de estos consorcios microbianos en el perfil de compuestos volátiles, y su influencia en el perfil sensorial. El objetivo de este trabajo es mostrar un panorama general de estos ecosistemas fermentativos desde el punto de vista biológico, matemático y tecnológico.

**Palabras clave:** Fermentaciones alcohólicas, vinos, aroma, cultivos mixtos, cultivos secuenciales, dinámica poblacional.

ABSTRACT

The population dynamics of mixed-culture and sequential-cultures in fruit-must fermentation is a very interesting problem from both the theoretical and technological stand point. By mixed-culture we refer to those fermentations in which more that one strain/species are present from the beginning of the process; and by sequential to those in which different microorganisms are added along the process. These kinds of fermentations play a key role in industry of fermented beverages. The mathematical models of such processes should represent the dynamics of multiple species growing in a mixture of substrates as the fruit must composition includes an important proportion of hexoses and pentoses. Since the flavor is a basic aspect of beverages, and in general of all fermented foods, it is fundamental to study the population dynamics and its impact in the organoleptic properties, through the influence in the volatile compound profile. The goal of this paper is to present a panorama of the investigations of these fermentative ecosystems from a biological, mathematical, and technological stand point.

INTRODUCCIÓN

Durante décadas la investigación en biotecnología se enfocó en cultivos puros, ahora los efectos sinérgicos en cultivos mixtos de microorganismos están siendo objeto de un creciente interés (Schink, 2002). La dinámica del cultivo mixto de microorganismos en mezclas de sustratos es un problema muy interesante desde el punto de vista teórico y tecnológico. Hoy en día se reconoce que para entender la presencia, el crecimiento y el papel que juegan los microorganismos en los bioprocesos se necesita un enfoque de ecología microbiana (Días & Wacher, 2003). Los cultivos mixtos que crecen en mezclas de sustratos juegan un papel fundamental en la bioingeniería.

La *producción de bioetanol*, es un ejemplo de mucho interés. La materia prima usada en la fermentación para producirlo consiste típicamente en una mezcla de hexosas y pentosas. Por tanto el bioetanol es el producto de la transformación de una mezcla de sustratos mediante microorganismos. Cuando un microorganismo crece en presencia de al menos dos sustratos, uno de ellos es normalmente agotado primero, lo que resulta en un cambio en la pendiente de la curva de crecimiento de biomasa. La población de levaduras usadas en la producción de bioetanol, a escala industrial, puede variar de acuerdo con las condiciones particulares del proceso de cada planta así como del estrés que el ambiente fermentativo impone a los consorcios de microorganismos. Por ello, las levaduras aisladas de una industria en particular pueden estar adaptadas a dichas condiciones y por tanto son mejores para ese proceso que una cepa comercial pura (da Silva-Filho *et al.*, 2005).

Los *alimentos fermentados* constituyen igualmente un área de gran interés. Su calidad y producción en condiciones controladas dependen del conocimiento y control de la microbiota presente (Días & Wacher, 2003). Un ejemplo clásico en alimentos fermentados con cultivos mixtos es el yogurt, el cual es el producto de la interacción de *Lactobacillus bulgaris* y *Streptococcus thermophilus* (Marshall, 1987). Para obtener una idea más detallada de la organización y dinámica de las comunidades microbianas en alimentos fermentados, será necesaria la detección de su actividad y el papel que esta juega en el producto terminado.

En este trabajo revisaremos tres aspectos fundamentales de las fermentaciones alcohólicas basadas en consorcios microbianos. Primero hablaremos del punto más importante en una bebida: el sabor, ya que éste determinará la aceptación del producto final. Después en la sección "los vinos y la dinámica poblacional" daremos un breve panorama del estado de arte de las fermentaciones alcohólicas que se llevan cabo con cultivos mixtos. Finalmente, en la sección "modelando la dinámica poblacional" presentaremos un panorama del modelado de estos complejos sistemas biológicos.

EL AROMA Y LOS VINOS

Según Abbott (1999) los atributos que más condicionan la aceptabilidad del alimento por parte del consumidor son los relacionados con la calidad sensorial u organoléptica, que incluye la apariencia, la textura, el aroma y el gusto. En este sentido, uno de los rasgos organolépticos más complejos y determinantes de la calidad sensorial es el aroma del alimento, que se puede definir como la sensación global producida por los compuestos que interaccionan con las terminaciones nerviosas sensitivas del gusto, del olfato y la visión (Goff & Klee, 2006). El aroma está compuesto por centenares de compuestos volátiles que pertenecen a distintas familias químicas y que se encuentran en muy variable concentración (Ruiz & Martínez, 1997). La elevada producción y la necesidad de encontrar alimentos aromáticamente estandarizados, requieren herramientas analíticas eficientes para la caracterización y algoritmos que permitan el control automático de la producción. Cabe mencionar que el umbral de percepción de las sustancias que condicionan el aroma puede variar desde µg/l a mg/l, pero no necesariamente por encontrarse en mayor concentración su incidencia será mayor (Riu, 2005). En este sentido, el impacto sensorial esta relacionado con las presencia de compuestos volátiles. Así pues, una de las principales variables a medir es la fracción aromática.

En enología lo usual es clasificar los aromas del vino en función de la etapa en la que se forman. El aroma puede provenir de la fruta, es el llamado *aroma primario* que incluye dos subcategorías: el *varietal* (compuestos volátiles libres presentes en la uva que dependerán de la variedad utilizada y sus características) y el *prefermentativo* (aromas que se liberan de su combinación con otras sustancias llamadas precursores, debido a la actividad enzimática provocada por la tecnología aplicada). El *aroma secundario* proviene de la levadura que se desarrolla durante la primera fermentación, está ligado a la presencia de ciertos tipos de enzimas y es el aroma mayoritario; y finalmente el *aroma terciario* o *post-fermentativo* es el que se forma durante la crianza. Este último se desarrolla mediante reacciones químicas y/o bioquímicas a partir de aromas de etapas anteriores (Riu, 2005).

El perfil aromático característico de las frutas depende de una mezcla compleja de compuestos químicos que se va formando durante la maduración del fruto, a través de distintas rutas bioquímicas a partir de precursores de las plantas (Gómez & Ledbetter, 1997; Lund & Bohlmann, 2006). El desarrollo de los aromas en las frutas tiene lugar durante el climaterio, que es el periodo crucial del proceso de maduración. Pequeñas cantidades de carbohidratos, lípidos, proteínas y aminoácidos se catabolizan y dan lugar a distintos compuestos volátiles. La velocidad de formación de estas sustancias aumenta después del inicio del climaterio y el proceso continúa tras la recolección de la fruta hasta que comienza la senescencia (Arthey & Ashurst, 1997). Al influir en la ecología del proceso de elaboración de vino, las levaduras contribuyen al sabor del vino. El metabolismo y la actividad enzimática de cada especie de microorganismo así como las combinaciones de éstas impactan en el aroma. Los característicos sabores frutales del vino se deben principalmente a la esterificación de los alcoholes superiores sintetizados por las levaduras.

## LOS VINOS Y LA DINÁMICA POBLACIONAL

La producción de *vino* es otro proceso de especial interés. Cabe mencionar que el término "*vino*" se aplica al líquido resultante de la fermentación alcohólica, total o parcial, del zumo de frutas, sin adición de ninguna sustancia. Se pueden encontrar además del vino de uva, vinos de mango y plátano entre otros vinos de frutas. Desde el punto de vista biotecnológico, el vino es el producto de complejas interacciones entre levaduras y bacterias que comienza desde que el fruto está en la plantación y continúa a lo largo de todo el proceso de fermentación, hasta el momento en que el producto es embotellado. En el proceso de elaboración de vino de uva, el tipo y cantidad de aroma depende de varios factores: microorganismos fermentantes, condiciones ambientales (suelo y clima), estado de la fruta, proceso de fermentación, pH del mosto, cantidad de dióxido de azufre, aminoácidos presentes en el mosto (Lilly *et al.*, 2000). En el caso de los vinos de uva está bien estudiado que aunque el tipo de uva y las condiciones de cultivo son claves en el sabor del vino, los microorganismos (en especial las levaduras) tiene un papel muy importante en las características organolépticas (Fleet, 2003). De León-Rodríguez *et al* (2006) reportaron que distintos tipos de mezcal joven, reposado y añejo producidos con mosto de *Agave salmiana*, presentaron diferente composición de compuestos volátiles como etanol, alcoholes superiores y ésteres, y que éstos contribuían de forma importante al perfil sensorial. De estos trabajos se puede concluir que la interacción entre la materia prima y el proceso de elaboración en la bebida son los responsables de las propiedades organolépticas de las bebidas alcohólicas provenientes de la fermentación.

Tradicionalmente la producción de vinos se ha realizado a partir de fermentaciones espontáneas de los mostos llevadas a cabo por cepas de levaduras endémicas residentes en las superficies de las uvas y de los equipos de las bodegas. Dentro de las fermentaciones alcohólicas basadas en consorcios microbianos las fermentaciones espontáneas se pueden considerar las pioneras de la biotecnología. Estas fermentaciones espontáneas de mostos son un complejo proceso que involucra la acción de diferentes géneros y especies de levaduras e incluso bacterias. El equilibrio entre los diferentes microorganismo presentes en la flora inicial, el orden de sucesión entre especies y la diversidad de la flora pueden variar entre un año y otro. Dando así origen a la diferencia en la velocidades de fermentación y a las características del vino de año a otro (Querol *et al.*, 1992 y 1994).

Existen argumentos a favor y en contra de las fermentaciones espontáneas. El principal argumento a favor indica que en estas fermentaciones se consiguen características organolépticas típicas de la zona que no estarían presentes si se utilizara un inóculo de cepas foráneas. Sin embargo la calidad del producto puede ser muy variable. La composición cualitativa y cuantitativa de las microbiota presente a lo largo de la fermentación del mosto puede depender principalmente de los siguientes factores: región de donde es originaria la fruta, procedimiento de producción, tipo de bebida a ser producida, concentración inicial de la microbiota, temperatura, pH, $SO_2$, y concentración de etanol (Torija *et al.*, 2001; Granchi *et al.*, 2002).

El uso de inóculos con poblaciones mixtas y/o inóculos secuenciales, constituye una herramienta importante para estandarizar el producto y preservar aquellas características deseables. Seleccionar cepas de una determinada región parece ser la solución para asegurar un producto estandarizado preservando las características organolépticas que distinguen a la zona de producción. Además del *Saccharomyces cerevisiae* se han reportado muchas otras levaduras presentes en la fermentación del vino: *Hanseniaspora guilliermondii*, *Kloeckera apiculata* (Romano et al., 1992, 1997a; Zironi *et al.*,

1993; Gil *et al.,* 1996), *Pichia anomala* (Rojas *et al*., 2001), *Candida stellata, Torulaspora delbrueckii* (Ciani & Maccarelli, 1998), *Candida valida, Bretanomyces bruxellensis, Rhodotorula aurantiaca, Deckera intermedia* (Mateo *et al*., 1991;) y *Candida catarellii* (Toro & Vázquez, 2002). Todas estas levaduras mejoran el *bouquet* del vino, pero no son capaces de terminar la fermentación debido a su poca tolerancia a altas concentraciones de etanol (Clemente-Jiménez *et al*., 2005). Por está razón, varios autores ya han estudiado fermentaciones usando mezclas de levaduras, ya sea inoculadas simultáneamente (Moreno *et al.* 1991; Gil *et al*., 1996; Erten, 2001) o de manera secuencial (Herraiz *et al.,* 1990; Zironi *et al.,* 1993; Toro & Vázquez, 2002). Cabe mencionar que estos trabajos no han sido abordados desde una perspectiva de sistemas dinámicos y no han generado modelos que permitan explorar aspectos de control y propiedades dinámicas. Los mecanismos de interacción de estos ecosistemas de la fermentación incluyen: producción de enzimas líticas, etanol, dióxido de azufre y efectos de tipo "*killer*"; competencia por los nutrientes, oxígeno, producción de dióxido de carbono.

MODELANDO LA DINÁMICA POBLACIONAL

Es muy importante la identificación y el entendimiento de las interacciones enológicas que ocurren en estos consorcios de levaduras y bacterias. Los trabajos pioneros de Jacob Monod y J. B. S. Haldane han servido como un punto de inicio de importantes modelos matemáticos que dan cuenta de diferentes aspectos del crecimiento microbiano en monocultivos por lote, lote-alimentado y continuos. Estos trabajos suponen cultivos puros. Actualmente existe una vasta cantidad de artículos sobre modelos matemáticos de crecimiento microbiano (Nielsen *et al*., 2003). Sin embargo, en el ámbito de cultivos mixtos la literatura se reduce considerablemente. Por supuesto, en general, cuando la complejidad de un problema crece, la posibilidad de analizarlo en términos precisos disminuye.

En el modelado de la mayoría de estos sistemas biológicos se asume de manera implícita que son de naturaleza continua, aplicando las ecuaciones diferenciales como la herramienta para su modelado. Las variables analizadas son denominadas estados, y son propiedades tales como la concentración de microorganismo (biomasa en g/l), concentración del producto (g/l), y concentración de sustrato (g/l); y en algunos casos concentración interna de enzimas. Sin embargo todos estos estados son muy difíciles de medir (sino es que imposible) en tiempo real. Se necesitan desarrollar modelos basados en ecuaciones diferenciales ordinarias que contemplen entre sus estados variables que sean fácilmente monitoreables en tiempo real. Dichos modelos permitirían el desarrollo de controladores que hagan frente a las perturbaciones inherentes a estos sistemas.

Se ha descrito como es posible inferir teóricamente la presencia de cultivos mixtos a partir de datos de biomasa total, usando una transformada ondeleta (Ibarra-Junquera *et al*., 2006[a]). La idea central desarrollada por Ibarra-Junquera *et al*., (2006[a]) se basa en el hecho de que todo evento a nivel metabólico (cambio de sustrato) o a nivel de interacciones entre especies (crecimiento mixto con competencia o sin ella) genera la presencia de singularidades en la señal de biomasa total. Es decir, estos eventos se asocian a la presencia de puntos en donde alguna derivada no exista, relacionando así el grado de singularidad a la naturaleza del evento que la provocó. La herramienta usada para detectar la existencia de singularidades y su grado fue la transformada ondeleta, la cual es un análogo de la transformada de Fourier a nivel local. En concreto, Ibarra-Junquera *et al.*, (2006[a]) mostraron teóricamente como cambios en la fuente de sustrato provocan singularidades en la segunda derivada de la señal de biomasa total mientras que crecimientos mixtos, sin competencia, inducen singularidades en la primera derivada.

Por otra parte se han realizado trabajados en el caso del mezcal donde se encontró que es posible monitorear en línea el proceso fermentativo a partir de medir el potencial redox durante la fermentación (Escalante-Minakata *et al.*, 2006[b]), asociando la señal de redox a la biomasa total. Ambas herramientas (el potencial redox y la transformada ondeleta) son de gran utilidad para entender la dinámica de las fermentaciones. Al estudiar la dinámica poblacional de las fermentaciones con cultivos mixtos se busca entender y eventualmente manipular (indirectamente) el mecanismo a través del cual una especie de microorganismo impacta a otra y finalmente a las propiedades organolépticas del producto de interés. Para incrementar la producción y la calidad del producto son necesarias técnicas para el monitoreo en línea y el control automático de estos complejos procesos fermentativos. Por otra parte, es importante resaltar que la mayoría de controles que operan actualmente en la industria de los bioprocesos están basados en modelos lineales del proceso, a pesar de que prácticamente todos los procesos biológicos son de naturaleza no lineal. Por ello es de esperarse que estrategias para su monitoreo y control basadas directamente en modelos no lineales muestren considerables ventajas y mejor desempeño en estos procesos fermentativos tan altamente no lineales.

Sin embargo, en la mayoría de los procesos bioquímicos es difícil desarrollar modelos matemáticos, basados en ecuaciones diferenciales ordinarias, que sean razonablemente precisos y cuyos valores estimados de parámetros sean confiables. No obstante, dichos modelos resultan fundamentales para la optimización y el desarrollo de estrategias de control del proceso. En particular en los reactores biológicos las incertidumbres en el modelo se deben a un limitado conocimiento del proceso real, a no linealidades, dinámicas no modeladas, presencia de ruido interno o externo, influencias del ambiente y parámetros variantes en el tiempo. La presencia de tales incertidumbres es la causa del desajuste entre el modelo optimizado y el proceso real, lo cual puede degradar el desempeño del controlador provocando serios problemas de estabilidad en el proceso. Por lo tanto resulta un reto de gran importancia, diseñar esquemas de control para procesos bioquímicos, que sean robustos a incertidumbres en el modelo.

Los primeros modelos de crecimiento mixto en un sustrato único, en quimiostato, mostraron teórica y experimentalmente, que no más de una especie sobrevive, sin importar la velocidad de dilución, ni la concentración de sustrato alimentado (Aris & Humphrey, 1977; Hansen & Hubell, 1980; Powell, 1958). Sin embargo, yacía una contradicción en este resultado, la llamada "paradoja del plancton" (Hunchinson, 1961). Por ello el problema de crecimiento mixto en una mezcla de sustratos atrajo el interés de muchos matemáticos. Este interés resultó en dos artículos fundamentales, que muestran que en presencia de múltiples sustratos limitantes, es importante especificar los requerimientos nutricionales satisfechos por los nutrientes (León & Tumpson, 1975; Tilman, 1977). Es decir, dos sustratos son mutuamente *sustituibles* si ambos satisfacen exactamente los mismos requerimientos nutricionales y con ello el crecimiento continúa incluso en ausencia de cualquiera de ellos. Entonces, podemos decir que dos sustratos son *complementarios* si satisfacen distintos requerimientos nutricionales y con ello el crecimiento es imposible en ausencia de cualquiera de ellos.

En años recientes han surgido modelos matemáticos que describen aspectos fisiológicos del crecimiento microbiano. Hanegraaf *et al.* (2000), desarrollaron un modelo que describe el mecanismo respiro-fermentativo de una levadura. El modelo toma en cuenta la presencia de múltiples rutas de asimilación de la fuente de carbono y su respuesta ante diferentes concentraciones de sustrato. Sin embargo, estos modelos solo son para cultivos puros, de ahí la importancia de desarrollar nuevos modelos que nos permitan mejorar y entender estos importantes fenómenos enológicos.

CONCLUSIONES

El mercado actual demanda no solo una calidad homogénea en los productos sino una alta calidad y un bajo precio. El estudio, el modelado y el análisis dinámico de los procesos de fermentación basados en cultivos mixtos permitirían no solo elucidar los mecanismos para influir en las propiedades organolépticas del producto final, sino también desarrollar estrategias para su control automático a nivel industrial. Estas herramientas conducirán al desarrollo de métodos sistemáticos de producción que permitan el desarrollo de productos homogéneos a lo largo de los años.